\documentclass[reprint,secnumarabic,amssymb,superscriptaddress,aps, prl, twocolumn]{revtex4-2} %nobibnotes,notitlepage,
\usepackage{amsmath,amssymb,bm}
\usepackage{graphicx}
\usepackage{color}
\usepackage[colorlinks=true,allcolors=blue]{hyperref}
\newcommand{\msd}{ \Delta r^2}

\newcommand{\beq}{\begin{eqnarray}}
\newcommand{\eeq}{\end{eqnarray}}
\newcommand{\mN}{\bm{\nabla}}
\newcommand{\br}{\bm{r}}

\newcommand{\krp}[1]{\textcolor{black}{#1}}

\DeclareUnicodeCharacter{2009}{\,}
\begin{document}
%\title{Non-Gaussian fluctuations in an enzyme cluster}
\title{Anomalous fluctuations in a droplet of chemically active colloids or enzymes}

\author{K. R. Prathyusha}
\email{equal contributions}
\affiliation{Max Planck Institute for Dynamics and Self-Organization (MPI-DS), D-37077 G\"ottingen, Germany}
\affiliation{Max Planck institute for Physics of Complex Systems, Dresden, Germany}
\affiliation {School of Chemical and Biomolecular Engineering, Georgia Institute of Technology, Atlanta, GA 30332, USA}
\email{current affiliation}
\author{ Suropriya Saha}
\email{equal contributions}
\affiliation{Max Planck Institute for Dynamics and Self-Organization (MPI-DS), D-37077 G\"ottingen, Germany}
\affiliation{Max Planck institute for Physics of Complex Systems, Dresden, Germany}
\author{Ramin Golestanian}%
\email{ramin.golestanian@ds.mpg.de}
\affiliation{Max Planck Institute for Dynamics and Self-Organization (MPI-DS), D-37077 G\"ottingen, Germany}
\affiliation{Max Planck institute for Physics of Complex Systems, Dresden, Germany}
\affiliation{Rudolf Peierls Centre for Theoretical Physics, University of Oxford, Oxford OX1 3PU, United Kingdom}
\date{\today}

\begin{abstract}
Chemically active colloids or enzymes cluster into dense droplets driven by their phoretic response to collectively generated chemical gradients. Employing Brownian dynamics simulation techniques, our study of the dynamics of such a chemically active droplet uncovers a rich variety of structures and dynamical properties, including the full range of fluid-like to solid-like behaviour, and non-Gaussian positional fluctuations. Our work sheds light on the complex dynamics of the active constituents of metabolic clusters, which are the main drivers of non-equilibrium activity in living systems.
\end{abstract}
\maketitle 

{\it{Introduction.---}}The non-equilibrium physical rules that determine the behaviour of active matter \cite{Gompper2020} can naturally be expected to provide clues towards unraveling the spatio-temporal self-organization observed in living systems. In particular, bio-chemical reactions facilitated by enzyme molecules and metabolic activity make the interior of a cell a non-equilibrium environment with persistent chemical gradients and fluxes \cite{cooper-thecell-2000, Testa2021Enzyme}. Theories of active phase separation, describing the phase behaviour of motile or living units, incorporate in addition to thermodynamic fluxes, particle currents stemming from non-equilibrium interactions, some examples of which are chemical interactions~\cite{ramin-PRL-12,Science-palacci-living-crystals-13,Stark_PhysRevLett.112.238303,MarenduzzoPRL,PRE-zwicker-frank-15,Rabea-Natphy-2017,Golestanian2016,jaime-ramin-19,bauermann2023formation}, quorum sensing~\cite{cates2015motility,QuorumBauerle2018}, non-reciprocity~\cite{Suropriya-PRX-20,Aparna-non-reciprocity-PNAS-20,Osat2023}, and catalysis~\cite{Matthew-PRL-2022}.  

Inside living cells, the structural compartmentalization of bio-molecules in the form of droplets are thought to help their function, such as regulating biochemical processes \cite{AnnuRevLLPS-Christoph-frank-14,Niebel2019}. These condensates are typically in a dynamic liquid-like state, although they can also exhibit solid-like properties when associated with pathological conditions~\cite{AAhyman_nat_review_disease-ageing-2021,BrangwynneScienceLLPS_physiology_health_17}. Due to the metastable nature of the liquid-like assemblies, they also exist in glassy or gel-like states that do not have the properties of a classical liquid~\cite{murray2017structure}. For instance, {\it in vitro} tracer diffusion measurements within phase-separated droplets have shown caging and other signatures of glassy behaviour \cite{LouiseScience2020}, while metabolic activity of bacteria has been shown to affect the diffusivity of the proteins within the cell cytoplasm \cite{parry2014bacterial,Bellotto2022}. 

Despite an overwhelming wealth of empirical observations, the interplay between enzymatic activity in the cytosol and the fluidity of protein condensates is still not understood from a mechanistic perspective. The non-equilibrium phoretic interactions, which naturally arise from chemical activity \cite{Golestanian2019phoretic}, have the potential to play a major role in such a regulation mechanism, in the same vein as the recently proposed mechanisms that may have led to the self-organization of metabolic cycles during the early stages of life formation \cite{OuazanReboul2023,VincentPRL,OuazanReboul2023NJP}.

\begin{figure}[b]
\vskip-0.5cm
%\begin{centre}
 \includegraphics[width= 1\linewidth]{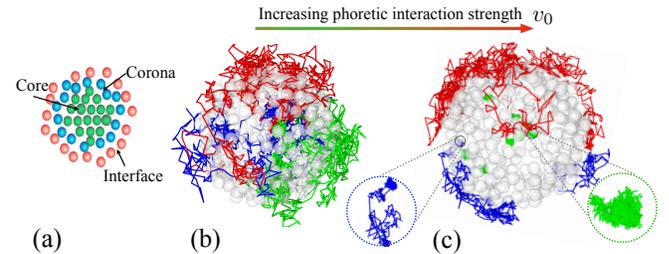}
%\end{centre}
\vskip-0.5cm
 \caption{(a) A schematic showing different regions---core, corona, interface---in a cross section of a cluster. (b) Trajectories of tagged colloids in a fluid-like cluster (weak phoretic interaction) are superimposed on transparent particles. A tagged particle travels freely throughout the entire cluster, see Supplemental Movie {SM1} \cite{SM}. (c) In an arrested cluster (strong phoretic interaction), particle tracks vary qualitatively depending on their initial location in the cluster. Completely trapped colloids (green track), and those that exhibit several cage breaking events (red and blue tracks) are seen in the same time window; see Supplemental Movie {SM2} \cite{SM}.}
\label{fig:schematic}
\end{figure}

\begin{figure*}
%\begin{centre}
 \includegraphics[width= 0.85\linewidth]{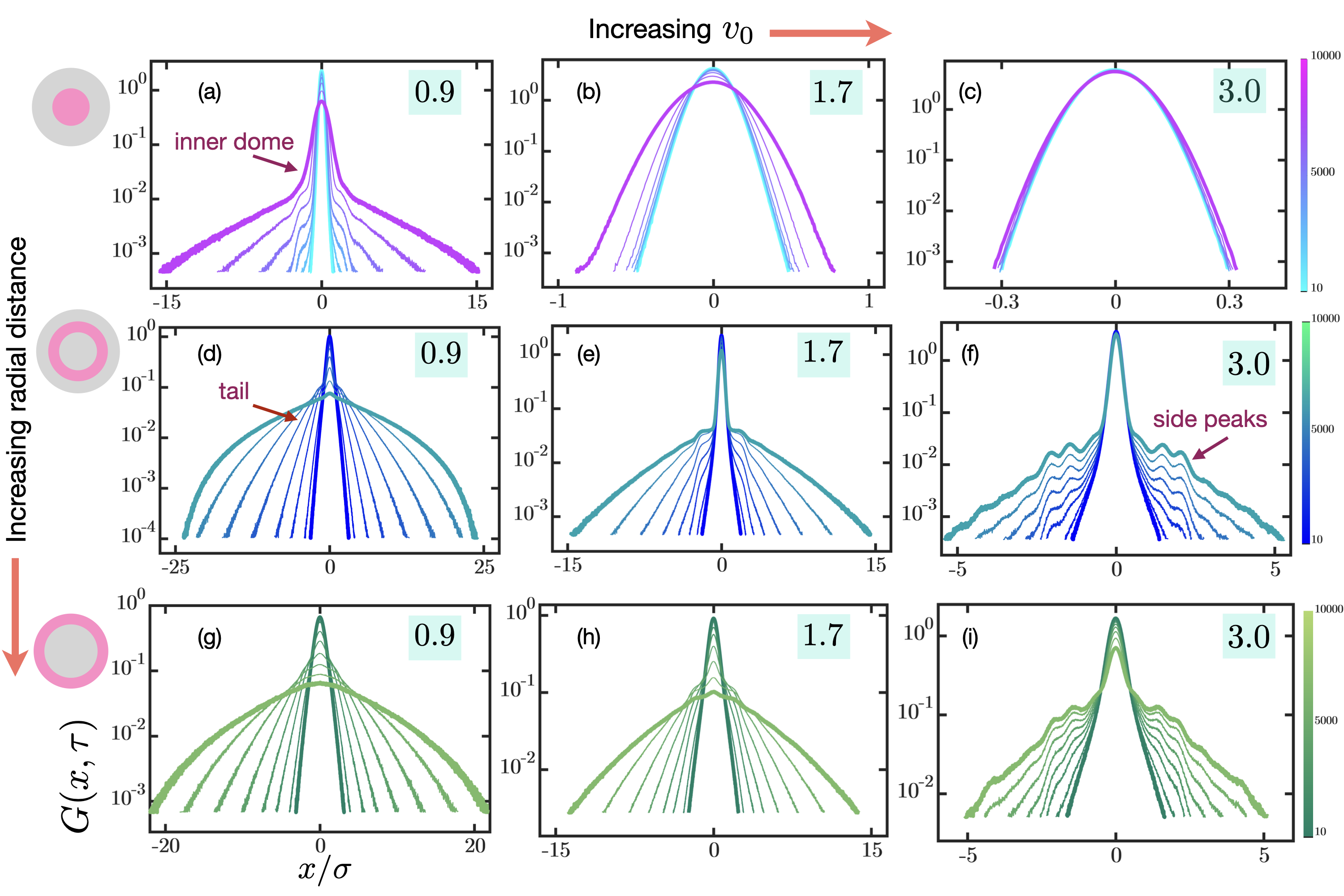}
%\end{centre}
\vskip-0.5cm
 \caption{Self-part of the Van Hove function (probability distribution of colloidal displacements). $G(x,\tau)$ in a partly arrested droplet is shown for a different values of $v_0$ (as indicated), increasing from left to right. $G$ for colloids initially located within a distance of $3-6 \sigma$ (core), $6-9 \sigma$ (corona) and $9-12 \sigma$ (interface) of the centre, are plotted in the three rows (from top to bottom) as indicated with illustrations. The colour-maps, which indicate the waiting time $\tau$, are chosen to be the same for colloids belonging to the same region in the cluster. A peak at $x=0$, indicative of a colloid trapped in a cage, is always present for colloids that are initially located in the core, as seen in panels (a-c). Such a peak is absent in fluids, and reveals slow relaxation. The peak is less pronounced for colloids originating in the corona, (see in panels d-f) and vanishes completely for those that start in the interface, (see  g-i). Those in panels (d-i) show pronounced tails extending until $\sim 10 \sigma$.  At $v_0 = 3.0$ fluctuations with magnitude of the order a few $\sigma$ show secondary peaks mirroring the structure formation, as seen in panels (f) and (i).}
\label{fig:svh}
\end{figure*}

\begin{figure}
%\begin{centre}
 \includegraphics[width= 0.85\linewidth]{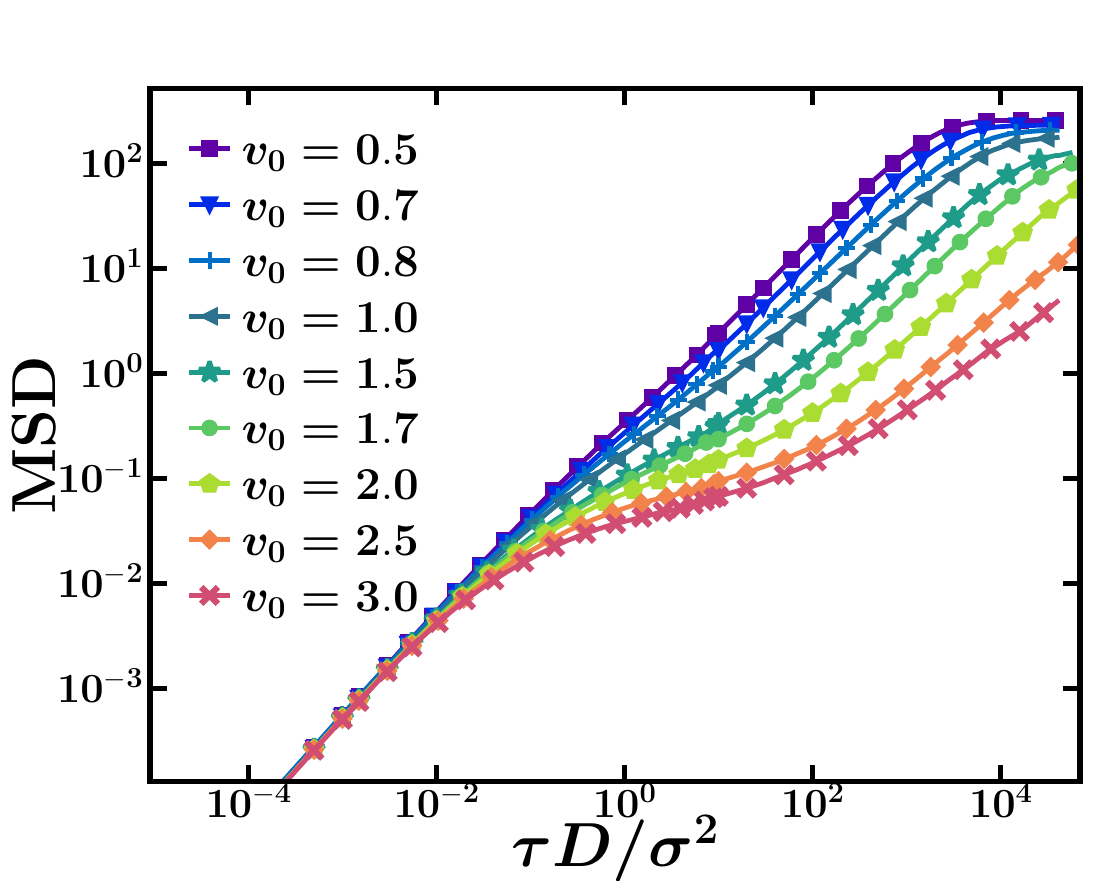}
%\end{centre}
\vskip-0.1cm
 \caption{Mean square displacement (MSD) for different values of $v_0$. A plateau develops in the arrested states as colloids are repeatedly trapped into, and released from temporary cages. The reduction in the MSD upon increasing $v_0$ is due to the increase in the attractive interaction between colloids, and saturation corresponds to the cluster size.} %It characterizes the average fluctuations in the motion of an individual colloid in the cluster. 
\label{fig:MSD}
\end{figure}

Here, we explore the complex dynamics within a droplet formed by long-ranged phoretic interaction between chemically active colloids or enzymes. Tracking the motion of a tagged bio-molecule provides information about the dynamics and structure inside a droplet~[see Figs.~\ref{fig:schematic} (a-c)]. The series of structural changes, which occur as the dimensionless coupling strength $v_0$ is increased, is accompanied by dramatic changes in the dynamics of a single particle (see Figs.~\ref{fig:svh} and \ref{fig:MSD}). The chemotactic collapse is cut off by steric repulsion, and the cluster undergoes a gradual transition from a fluid-like state (see Supplemental Movie { SM3} \cite{SM}) to a solid-like state~(see Supplemental Movie {SM7} \cite{SM}) as $v_0$ is increased. At intermediate values of $v_0$ (see Supplemental Movies SM4-6 \cite{SM}), the cluster develops a solid core surrounded by a relatively mobile region that we call the corona, which in turn is followed by an interface consisting of chemically active colloids that are nearly free [see Fig.~\ref{fig:schematic}(a)]. Using the distribution of the positional fluctuations calculated as a function of the step size and initial location of the colloid as our main tool, we probe the glassy dynamics in this mesoscopic droplet and provide several experimentally testable results.

{\it{Theoretical model.---}}We consider $N$ colloids of radius $\sigma$ within a spherical container of radius $R$, with the stochastic trajectory of the $i$th particle ($i=1,\dots,N$) denoted as $\br_i(t)$. Each active colloid catalyzes a chemical reaction converting a reactant, assumed to be abundantly available, into a product at a rate $\alpha$. They generate a chemical field $c(\br,t)$ at position ${\br}$ % measured from the origin of the confining container, 
that evolves following the diffusion equation with sources at $\bm{r}_i$, namely, 
 \begin{eqnarray} 
 \partial_t c - D_c \nabla^2 c = \alpha \sum_i \delta(\br - \br_i), 
\label{eq:evolutionC} 
 \end{eqnarray}
 where $D_c$ is the diffusion coefficient of the chemicals. Imposing the boundary condition $c(|\br| = R,t) = 0$ ensures that the chemicals are continuously generated in the container and extracted at the boundary, hence creating a non-equilibrium steady-state. Variation of $c$ on the colloid surface establishes a diffusiophoretic slip velocity and thus net drift with a velocity $-\mu \bm{\nabla} c$, where $\mu$ is the diffusiophoretic mobility (that is negative for attractive phoretic interactions) \cite{Golestanian2019phoretic}. The equation of motion of the $i$th colloid is given as
 \begin{eqnarray}
 \dot{\bm{r}}_i = -\mu \bm{\nabla} c(\bm{r}_i,t) + \sum_{i \neq j} {v}({r}_{ij})\hat{\bm{r}}_{ij} +  \bm {\zeta }_i, 
 \label{eq:evolutionPos}
 \end{eqnarray} 
 where $\bm{r}_{ij} = \br_i - \br_j$, $r_{ij}=|\bm{r}_{ij}|$ and $\hat{\bm{r}}_{ij} = \bm{r}_{ij} /r_{ij}$. ${v}(r_{ij})=  24\epsilon [ 2{(2\sigma)^{12}}{r_{ij}^{-13}}  - {(2\sigma)^6}{r_{ij}^{-7}}] $ is a derivative of the  Weeks-Chandler-Anderson potential \cite{weeks-chandler-anderson-jcp-71}. It imposes steric repulsion between the colloids and vanishes for $r_{ij}>2^{1/6}(2\sigma$). The parameter $\epsilon$ combines the strength of repulsion and the viscous damping and is kept constant at unity. The random fluctuations are included through the white noise term, $\bm {\zeta }$, with zero mean and intensity $2D$, where $D$ is the thermal diffusivity of the colloids.
 
 Assuming a separation of scale between the sizes of the chemicals and the colloids, we can use the quasi-stationary solution of Eq. (\ref{eq:evolutionC}) since $D \ll D_c$. Moreover, we use the far-field approximation~\cite{soto-ramin-PRL-14,jaime-ramin-19} and ignore corrections due to the proximity of colloids. This approximation is justified since exact solutions have shown that near-field effects are unimportant for exactly similar active colloids~\cite{BabakPhysRevLett.124.168003}. With these approximations, the chemical gradient $\bm{\nabla}c$ can be determined explicitly as a function of colloid positions $\br_i$ as follows
\begin{eqnarray}
\bm{\nabla} c(\br_i,t) =\frac{\alpha}{4\pi D_c} \left[\sum_{j \ne i}^N  \frac{\br_{ij}}{r_{ij}^3}-\sum_{j = 1}^N\frac{(R/r_j){{\br}_{ij}^\prime}}{r^{\prime 3}_{ij}} \right],
 \label{eq:poisson_solution3}
\end{eqnarray} 
Where $\br^{\prime}_{ij}=\br_i-\frac{R^2}{r_j^2}\br_j$, $r_{ij}^\prime=|\br^{\prime}_{ij}|$  and $r_j=|\br_j|$.
Scaling position by $\sigma$ and time by $\sigma^2 D^{-1} $, we identify a dimensionless constant $v_0 = |\mu| \alpha/(D D_c \sigma)$ which determines the strength of the interactions with respect to the fluctuations. 

Particle trajectories are obtained by the Euler integration of \eqref{eq:evolutionPos} with a time step $\Delta t=0.001 $. The data presented in this paper is for $N = 1000$ unless otherwise specified. The value of $v_0$ varies between $0.5$ and $5.0$. The colloids assemble to form a single spherical droplet of size equal to a few colloidal radii ($\sim 10\sigma$) at the centre of the confining sphere due to the long-range interaction between them. 

{\it{Self-part of the Van Hove functions.---}}The central result of our work is the analysis of anomalous fluctuations to illustrate several aspects of the dynamics that follow from the chemically mediated long-range interactions. We do so by calculating the self-part of the Van-Hove functions (SVH)~\cite{VanHove1954}. In a fluid with no internal structure \cite{Theory-Of-Fluids}, the SVH is Gaussian, while in a supercooled fluid it is Gaussian with exponential tails \cite{Pinaki-2007, colloidalglass-weitz-Science2000}. We calculate the SVH by {\it{distinguishing}} the initial position of the colloid in the droplet. As a result the distribution depends on whether the colloids were located initially in the frozen core or the corona. The self part of the Van-Hove function $G(x,\tau)$ is defined as follows  
\beq
G( x,\tau) = \frac{1}{n} \sum_{i = 1}^{n}\langle \delta[x - (x_i(\tau+t) - x_i(t) ] \rangle.
\label{eq:SelfVanHove}
\eeq
$G(x,\tau)$ is the probability distribution function that a colloid traverses a displacement $x$ in an interval of time $\tau$. Variation of $G(x,\tau)$ with the waiting time $\tau$ provides information about the changing neighbourhood of a colloid. The $\tau$ dependent step-size is thus simply the distance $x_i(t+\tau) - x_i(t) $, where the time $t$ is chosen large enough such that the cluster has reached a steady-state. In the definition \eqref{eq:SelfVanHove}, the index $i$ is used to average over a total of $n$ number of colloids which are at time $t$ located in a particular shell from the centre of mass of the cluster (as shown schematically in Fig. \ref{fig:svh}). Since $G$ is identical for fluctuations in the three orthogonal directions, we present an average over all three directions.

We find that $G$ reveals a wealth of information about the spatial dependence of structural rearrangements within the cluster when it is calculated for those located in the core, the corona, or the interface.  For small $v_0$, $G$ is well approximated by a Gaussian irrespective of the initial location of the colloids, and its width increases as $\sim \sqrt{\tau}$ for all $\tau$. For a value of $v_0$ for which the cluster is close to the solid state, $G$ begins to show signatures of trapping at small $\tau$, and cage-breaking dynamics at larger values of $\tau$. Calculated for colloids in the core with initial positions within $3-6 \sigma$ from the centre of the cluster, $G$ shows a sharp peak at $x=0$ as seen in Figs. \ref{fig:svh}(a-c). The uni-modal graph falls sharply within $x \sim \sigma$ showing that the colloids in the interior of the cluster for $v_0 = 1.7$ are caged by their neighbours [see Figs. \ref{fig:svh}(b-c)]. For colloids initially located in the corona within $6-9 \sigma$ from the centre, the distribution clearly develops a tail - which broadens with increasing $\tau$ [see Figs. \ref{fig:svh}(d-f)]. For colloids with initial locations in the interface within $9-12 \sigma$, $G$ shows tails whose widths increase with increasing $\tau$ [Figs. \ref{fig:svh}(g-i)]. Thus the dynamics varies greatly from the centre to the surface - the innermost colloids vibrate in nearly permanent cages (over the time-scale of the simulations), the ones which are in the corona are trapped for variable duration of time and released before they are trapped again. Colloids at the interface typically make long excursions [Figs. \ref{fig:svh}(g-i)]. In the fully arrested state, $G$ shows a single peak at $x=0$ of nearly constant width for all $\tau$ [Figs. \ref{fig:svh}(c), (f) and (i)]. Deep into the solid state fluctuations close to $x=0$ develop additional features such as secondary peaks that reflect the underlying positional order [Figs. \ref{fig:svh}(f) and \ref{fig:svh}(i)]. Such side-peaks have been reported in other active matter systems due to the action of molecular motors in a gel \cite{NirGov-PRE}.

To elucidate the non-Gaussian nature of the fluctuations, we fit the inner dome %[see Fig. \ref{fig:svh}(a) and SM~\cite{SM}, Fig.~S4(a)] 
and the outer tail of $G$ %[see Fig. \ref{fig:svh}(d) and SM~\cite{SM}, Fig.~S4(b)] 
to a family of curves called the $q$-Gaussian~\cite{Tsallis1988,Tsallis_PhysRevLett.84.2770}, 
which provides a framework to describe systems with long-range interactions ~\cite{tsallis2023review} (see SM for details \cite{SM}). The $q$-Gaussian of a length $x$ scaled by $\ell$ is $\left[ 1 - (1-q) (x/\ell)^{2} \right]^{\frac{1}{1-q}}$. In general, the range of the exponent $q$ is  $-\infty<q<3$, approaching the Gaussian as $q\to 1$. For $q<1$, the domain of the function is bounded, i.e. $-1 <x/\ell<1 $, while for $q>1$, $x/\ell$ is unbounded. $G$ calculated for colloids in the core is well fitted by the $q = 1.45$, % (see SM~\cite{SM} Fig.~S4(c)), 
corresponding to strong correlations ~\cite{tirnakli2022entropic}. For $G$ calculated for those in the corona, the tail is fitted by $0<q < 1$. A fitted value of $q$ smaller than unity suggests that fluctuations smaller than a length scale are suppressed. We also calculate the distribution of time intervals between successive cage-breaking events. %(see SM~\cite{SM} Fig S5(a)). 
The distribution changes from being exponential in the fluid state to a power law in the solid (see SM for details \cite{SM}). % (see SM~\cite{SM} Fig S5(b)). 

{\it{Mean square displacement.---}}We  discuss fluctuations in colloid position as captured by calculating the mean square displacement (MSD) for all colloids in the droplet (see SM for MSD calculated by distinguishing the initial location of the colloids~\cite{SM}). The centre of mass of the mesoscopic cluster diffuses while  it also rotates as a whole. To measure relative displacements of particles, we transform to a body-fixed frame of reference located at the centre of the cluster using methods described in \cite {hunter-weeks-optics-11}. The MSD is calculated by averaging over trajectories of all particles as $ \mbox{MSD}(t)= N^{-1} \sum_{i=1}^{N}\langle |{\bf r}_i(t)-{\bf r}_i(0)|^2 \rangle$, %, where the angular brackets denote an ensemble average. 
and shown in Fig. \ref{fig:MSD}, for $v_0$ in the range $0.5-3.0$. A plateau, defined as a flattening of the MSD curve after an initial diffusive regime is visible at sufficiently large values of $v_0$, and is particularly prominent in the arrested state. It  emerges as the colloids get trapped or caged by their neighbours and spend a long time inside those cages. Such a sub-diffusive plateau is a signature of the motion of active particles in complex and crowded environment~\cite{bechinger2016active}, in contrast to a freely moving active particle. The trapped colloids escape from their cages after a time scale that increases with increasing $v_0$, as apparent in Fig. \ref{fig:MSD}. Note that the trapping occurs at a comparable time scales ($ \sim \tau$) in all $v_0$s but in contrast, escape from a cage is a collective manoeuvre and the related timescale, increases over three orders of magnitude from the fluid to the solid phase. %~(see SM~\cite{SM} Fig.~S2). 

\begin{figure}
%\begin{centre}
 \includegraphics[width= 0.85\linewidth]{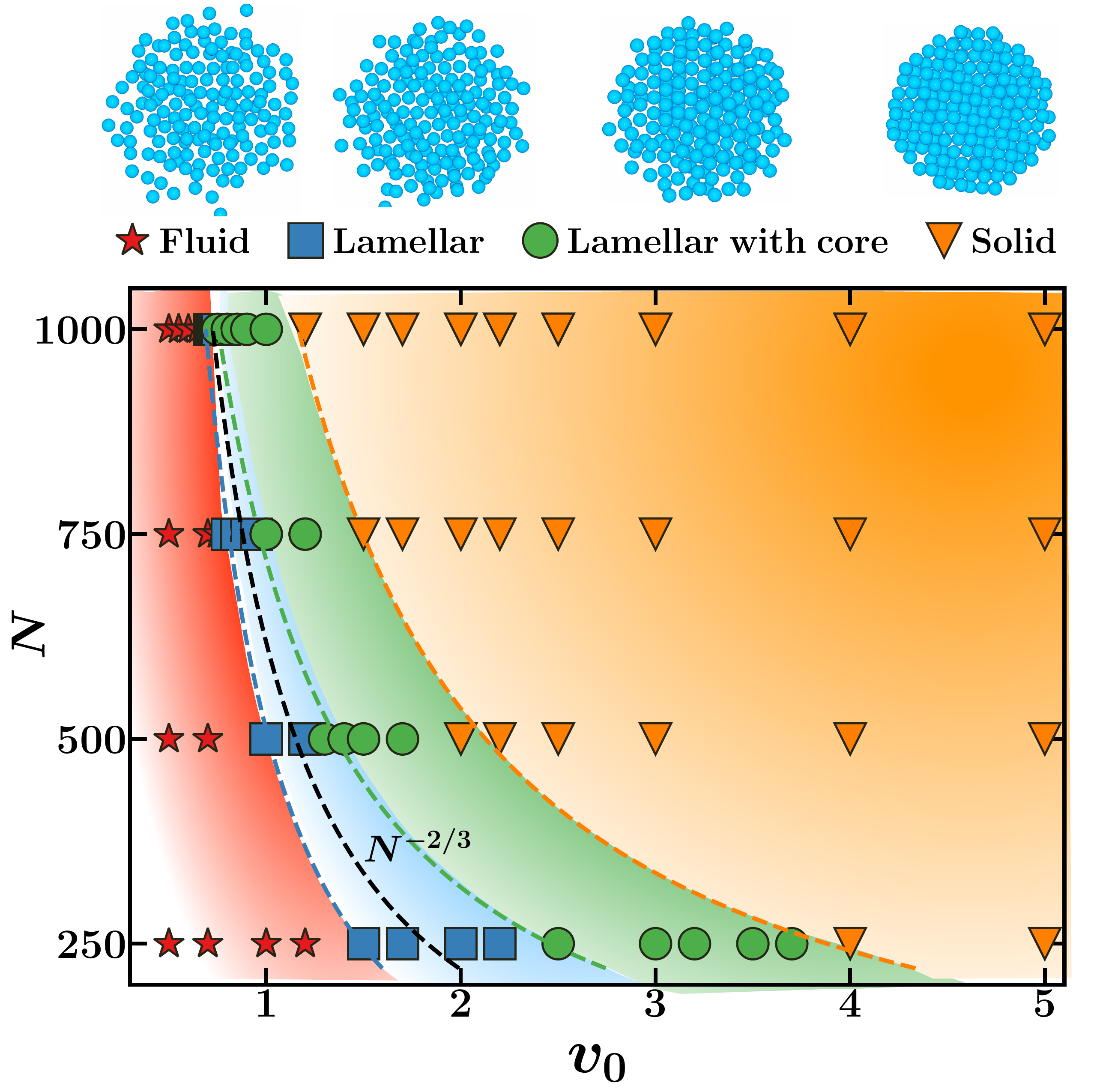}
%\end{centre}
\vskip-0.1cm
\caption{State diagram in $N-v_0$ plane, showing fluid, lamellar, lamellar with core and arrested states. Note that the phase boundaries shift towards smaller values of $v_0$ with increasing system size. The dashed black line is the power law scaling expected using a dimensional analysis described in the text. Snapshots from simulations illustrating typical structure of the droplet in each state are shown at the top. %A thin slice ($\sim 3 \sigma$) of the droplet is shown for clarity.
} %State diagram. 
\label{fig:states}
\end{figure}

{{\it{State diagram.---}}We construct a state diagram by varying both $v_0$ and the total number of particles. Using cues from both the arrangement of colloids within the droplet and their dynamics, we identify two stages in between the fluid-like and the solid-like droplet which are called `lamellar' and `lamellar with core' (see Fig.~\ref{fig:states}). Periodic deviations around the smooth radial density profile, % [see SM~\cite{SM} Fig. S1(a-d)], 
calculated from the centre of mass of the droplet, serves as the metric to distinguish a {\it{lamellar droplet}} that has developed shells like an onion %[see SM~\cite{SM} Fig. S1(b), $v_0=0.7$]  
from a fluid-like droplet without spatial ordering (see SM for details \cite{SM}). %[see SM~\cite{SM} Fig.~S1(a), $v_0=0.5$]. 
The dynamics in a lamellar droplet is still fluid-like (Supplemental Movie {SM4} \cite{SM}). Development of structural inhomogeneities in a similar density of the positions of colloids within a spherical shell at an even higher $v_0$ signals the transformation of a lamellar droplet into a lamellar droplet with a core. % [SM \cite{SM} Fig. S1(g)]. 
In the state `lamellar with core', the droplet develops an inner core within which colloids are immobile due to the surrounding dynamic layer that we call the corona (see Supplemental Movies {SM5} and {SM6} \cite{SM}). Irregular sharp peaks in densities %[see SM~\cite{SM} Figs. S1(d),(h)] 
reveal that the solid structure formed is not isotropic due to the finite size of the cluster (see SM for details and  Supplemental Movie {SM7} \cite{SM}). Figure \ref{fig:states} displays a state diagram in the $N-v_0$ plane and shows that the state boundaries shift to lower values of $v_0$ with increasing $N$; we argue that this is a feature of the long-range interactions mediated by the chemical field. Ignoring fluctuations of colloids at the interface, it is reasonable to assume that at the boundary of the fluid-solid transition, the cluster size scales with $N$ as $L^3 \sim \sigma^3 N$; using this length scale we find that gradients of $c$ scale as $\alpha N/(D_c L^2)$. Balancing the chemotactic drift $-\mu \mN c P$ with the diffusive current $-D \mN P$ in the Fokker-Planck equation for the probability $P$, we find $v_0 \sim (\sigma/L)^2 \sim N^{-2/3}$~\cite{ramin-PRL-12}.

{\it{Concluding remarks.---}}The strength of the effective interaction mediated by the collective response of the colloids to the chemical field determines whether the cluster resembles a fluid or is rather in a hybrid state with a central core resembling a solid and an outer corona of relatively freely moving colloids. We observe strikingly different dynamics in the two cases: in the first case, the colloids are free to span the full cluster while the motion of the caged ones is restricted to a fraction of the colloids in the second case. We observed narrow Gaussian peaks and extended tails in the distributions of particle displacements, similar to what is ubiquitously observed in glassy systems \cite{Pinaki-2007, Kegel2000,  colloidalglass-weitz-Science2000}, where the dominant dynamics of the particles is random hopping and trapping, also including solid-liquid interfaces \cite{Dschwartz-sl-interface-13}. The crucial difference between the system at hand and the classically studied systems is that in our system both the inner dome and the tail are fitted by the $q$-Gaussian. Our analysis of the fluctuations by using the $q$-Gaussian distribution thus elucidates experimentally testable differences between jammed or super-cooled passive matter and those in chemically active matter.

We show that non-equilibrium interactions driven by the phoretic response of chemically active enzymes or colloids to collectively generated gradients can lead to an effective mechanism of regulation for the structural and mechanical properties of metabolically active protein condensates. The regulation is achieved in a seemingly counter-intuitive sense, as stronger catalytic fluxes lead to solidification of the core of the cluster while allowing the particles at the interface to move more efficiently. This can be understood as being due to the balance between the short-range repulsion that arises from equilibrium interactions and long-range attraction that arises from the non-equilibrium chemical activity. We note that the opposite sense of regulation can also be envisaged if we implemented attractive equilibrium interactions and repulsive non-equilibrium interactions. Our results thus provide a mechanistic understanding of how such naturally occurring non-equilibrium interactions can contribute to the active regulation of the structural and dynamical properties of intra-cellular condensates.  

\begin{acknowledgements}
KRP and SS contributed equally to this work. We thank J. Agudo-Canalejo, A. Amiri, B. Mahault, Y. Pollack, S. Rulands, A. Vilfan, F. Ziebert, and D. Zwicker for useful discussions. RG acknowledges Martin Gutzwiller Fellowship of the Max Planck Institute for the Physics of Complex Systems in Dresden. This work has received support from the Max Planck School Matter to Life and the MaxSyn-Bio Consortium, which are jointly funded by the Federal Ministry of Education and Research (BMBF) of Germany, and the Max Planck Society. We acknowledge the use of the Max Planck Computing and Data Facility (MPCDF) in Garching.   \end{acknowledgements}
%\bibliography{Enzymecluster}{}
%

%\ifarXiv
   % \foreach \x in {1,...,\numbersupplementpages}
    %{
      %  \clearpage
        %\includepdf[pages={\x,{}}]{\SM_Cluster.pdf .pdf}
    %}
%\fi

%\documentclass[secnumarabic,amssymb,superscriptaddress,aps]{revtex4-2} %nobibnotes,notitlepage,
%\usepackage{amsmath,amssymb,bm}
%\usepackage{tocloft}
%\usepackage{ulem}
%\usepackage{graphicx}
%\usepackage{color}
%\usepackage[colorlinks=true,allcolors=blue]{hyperref}%
%\newcommand{\msd}{ \Delta r^2}
%\newcommand{\classoption}[1]{\texttt{#1}}
%\newcommand{\macro}[1]{\texttt{\textbackslash#1}}
%\newcommand{\m}[1]{\macro{#1}}
%\newcommand{\beq}{\begin{eqnarray}}
%\newcommand{\eeq}{\end{eqnarray}}
%\newcommand{\mN}{\bm{\nabla}}
%\newcommand{\br}{\bm{r}}
%\newcommand{\bn}{\hat{\bm{r}}}
%\newcommand{\moo}{\mbox{\tiny{out}}}
%\newcommand{\moi}{\mbox{\tiny{in}}}
%\newcommand{\env}[1]{\texttt{#1}}
%%\setlength{\textheight}{9.5in}
%\newcommand{\Suro}[1]{\textcolor{blue}{#1}}
%\newcommand{\krp}[1]{\textcolor{black}{#1}}
%\newcommand{\krpnote}[1]{\textcolor{magenta}{#1}}
%\newcommand{\ssnote}[1]{\textcolor{blue}{#1}}
%%\usepackage{xr}
%\usepackage{cleveref}
%\externaldocument[supp-]{SI_PRL}

%\DeclareUnicodeCharacter{2009}{\,}
\newpage
\widetext
\makeatletter
\renewcommand{\fnum@figure}{\figurename~S\thefigure}
\makeatother
\begin{center}
\textbf{\large {Anomalous fluctuations in a droplet of chemically active colloids or enzymes  \\
{\it Supplemental Material}}}
\end{center}
\setcounter{figure}{0}

\renewcommand{\theequation}{S\arabic{equation}}
%\renewcommand{\thefigure}{S\arabic{figure}}
%\begin{document}
%\title{Non-Gaussian fluctuations in an enzyme cluster}
%\title{Anomalous fluctuations in a droplet of chemically active colloids or enzymes  \\
{%\it Supplemental Material}}

%\author{K. R. Prathyusha}
%\email{equal contributions}
%\affiliation{Max Planck Institute for Dynamics and Self-Organization (MPI-DS), D-37077 G\"ottingen, Germany}
%\affiliation{Max Planck institute for Physics of Complex Systems, Dresden, Germany}
%\affiliation{School of Chemical and Biomolecular Engineering, Georgia Institute of Technology, Atlanta, GA 30332, USA}
%\email{current affiliation}
%\author{ Suropriya Saha}
%\email{equal contributions}
%\affiliation{Max Planck Institute for Dynamics and Self-Organization (MPI-DS), D-37077 G\"ottingen, Germany}
%\affiliation{Max Planck institute for Physics of Complex Systems, Dresden, Germany}
%\author{Ramin Golestanian}%
%\email{ramin.golestanian@ds.mpg.de}
%\affiliation{Max Planck Institute for Dynamics and Self-Organization (MPI-DS), D-37077 G\"ottingen, Germany}
%\affiliation{Max Planck institute for Physics of Complex Systems, Dresden, Germany}
%\affiliation{Rudolf Peierls Centre for Theoretical Physics, University of Oxford, Oxford OX1 3PU, United Kingdom}
%\date{\today}

%\maketitle

%\tableofcontents

%\vskip1cm

%\begin{document}
%\section{}
 
%We have not included fluid flows in our simple model for an enzyme cluster. 
%We discuss here, how the effect of fluid flows can be incorporated. the diffusiophoretic interactions contribute stresses equal to $\rho \mathbf{E}$. 
\vspace{0.5cm}
In this supplement we discuss the details of the simulations used to study a system of phoretic colloids. We discuss the various measures used in the main paper to quantify the different phases. We also present further results on the dynamics such as the local slope of the MSD showing the clear existence of a plateau, the MSD for colloids originating in different layers, and finally the distributions for the time periods that the colloids remains trapped in a cage (the waiting time distribution).

\section { Simulation details} 
We place colloids randomly inside a spherical container of radius $50\sigma$. We do an initial run with repulsive interactions alone for further homogeneous mixing of colloids. We employ an additional reflective boundary condition to restrict the movement of colloids outside the container during this process. This homegenous configuration is used as initial conifiguration for the study of phoretically interacting colloids.  Each of these colloids  produce a spherically symmetric chemical field. Then  we allow the system to relax for $10^4\tau$ followed by a data production run of $10^6\tau$. We choose the sign of $\mu$ (Eqn. (2) in the main article) such that phoretic interaction is attractive and the colloids assemble to form a single spherical droplet at the center of the confining sphere. The boundary condition for the concentration field at the the edge of the confining sphere being zero ensures that the droplet is formed at the center of the container. As we are interested in the properties of a dense cluster, the effect of hydrodynamics is neglected against the other dominant interactions present in the system. We vary the number of colloids in our simulations, $N = 250, 500, 750, 1000$. We set the radius of the colloids $\sigma$ as the unit of length and unit of time as $\sigma^2/D$ where $D$ is the diffusion coefficient of colloids and strength of fluctuation. The strength of interactions is determined by a dimensionless number $v_0 = \mu \alpha D^{-1} D_c^{-1} \sigma^{-1}$, where $D_c$ is the diffusivity of the chemical field and $\mu$ is the diffusiophoretic mobility. In this work $v_0$ is varied between $0.5$ to $5.0$. 

%\clearpage

\begin{figure}[!h]
\begin{center}
    \includegraphics[width=0.7\textwidth]{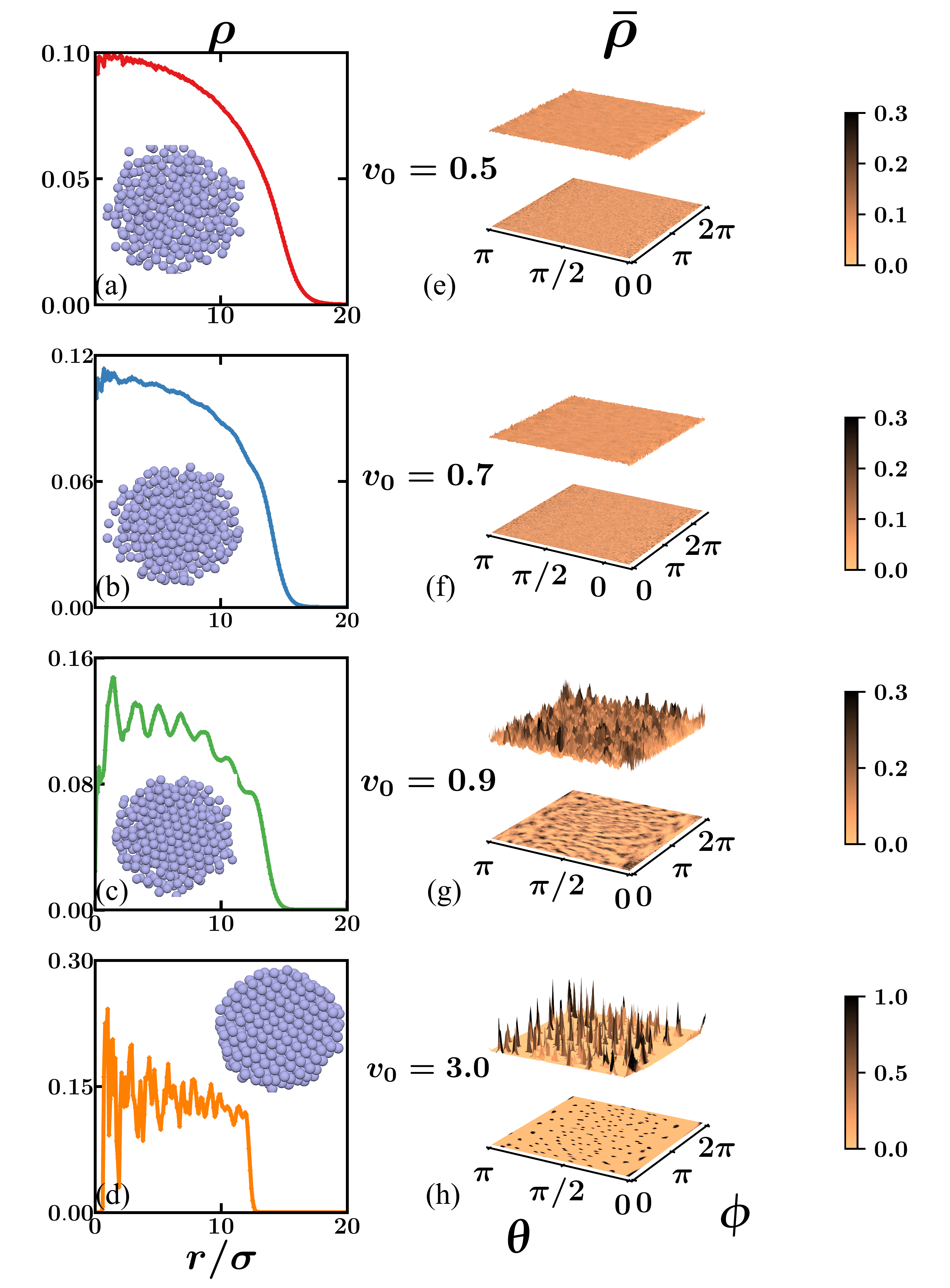}
    \end{center}
    \caption{Structural arrest with increasing strength of interaction. (Left) Radial density profile $\rho(r)$ as a function of the distance $r$ from the center of the cluster for four representative values of $v_0$ in the (a) fluid  ($v_0=0.5$), (b) lamellar ($v_0=0.7$), (c) lamellar with core ($v_0=0.9$) and (d) solid($v_0=3.0$) phase. Corresponding values of $v_0$ are indicated in each sub-figure. The smoothly decaying tail of $\rho$ in fluid-like cluster changes to an abrupt decrease near the edge of a solid cluster; while in the lamellar state a layering develops with periodicity equal to the colloid diameter. Inset: a snapshot showing all colloid positions in the cluster at a random instant of time. (Right) Distribution $\bar\rho(\theta, \phi)$ of the angular position ${\bf r}/r$ of colloids is shown in the four states for the same values of $v_0$ as in (Left) (e)     fluid phase ($v_0=0.5$), (f) lamellar phase ($v_0=0.7$), (g) lamellar with core ($v_0=0.9$) and (h) solid~($v_0=3.0$); both a three dimensional snapshot and a two dimensional projection are shown. Inhomogeneities apparent in the transition state transform to sharp well defined peaks in arrested state, showing that particles hold their positions. }
     \label{sifig:rho_rhobar}
  \end{figure}

%\clearpage
  
\section{Radial density profile inside the colloidal droplet }
The colloids assemble to form a single spherical cluster of radius $\approx 10\sigma$ at the center of the confining sphere due to attractive long-range interaction between colloids and boundary condition of $c$. The qualitative changes that we observe in the dynamics as $v_0$ is tuned, is linked to the development of structure in the cluster from liquid-like to a solid state.

For droplets of all sizes, the transition occurs in a two steps \krp{as $v_0$ is increased}. In the first step the droplet transitions from a fluid like loosely bound structure to one that has concentric spherical layers like an onion   and we call this state the lamellar phase. In a fluid-like cluster, the radial density profile $\rho(r)$, measured from the centre of mass of the cluster, is constant near the centre and decays smoothly zero near the periphery of the cluster (see Fig.~\ref{sifig:rho_rhobar}(a), $v_0=0.5$). 
In the lamellar phase the particles organise themselves in concentric spherical shells as visible in the undulations that develop in $\rho (r)$(see Fig.~\ref{sifig:rho_rhobar}(b), $v_0=0.7$). The radial density $\rho(r)$, obtained by averaging over spherical shells, clearly shows the transition from a fluid to a "lamellar structure" (layered like an onion) (See   Fig.~\ref{sifig:rho_rhobar}(b), $v_0=0.7$)). The value at which it develops undulations with a length scale $ \approx 2 \sigma$ is used to determine the lowest boundary of the state diagram (Main article Fig. 4) that denotes the transition from the fluid-like cluster to a lamellar structure.

We also measure the joint distribution $\bar \rho({\theta, \phi})$  of the polar and the azimuthal angle of the angular position $\hat n = {\bf r}/r$ to identify the dynamics of the colloids inside a spherical layer. Even though particles arrange shell like structure, the dynamics of the colloids is similar to fluid phase, see the homogeneous distribution $\bar \rho({\theta, \phi})$  both in Fig.~\ref{sifig:rho_rhobar}(e) and (f).

The lamellar phase develops an inner  core as $v_0$ is increased further, within which colloids are immobile due to the surrounding dynamic layers which we call corona. Colloids are trapped for increasingly longer time within by their neighbours as we go radially inward from the periphery of the droplet and we call the state  "lamellar phase with core" (see Fig.~\ref{sifig:rho_rhobar}(c)). Even though in lamellar structure, the exchange of neighbouring colloids is similar to what is seen in a fluid phase, lamellar with core differ in dynamics.   
In lamellar with core phase, $\bar \rho({\theta, \phi})$ shows inhomogeneities due to the particle arrest at the core (\ref{sifig:rho_rhobar}(g) $v_0 = 0.9$). Finally the entire cluster, with the exception of the colloids right at the interface, are arrested. (see ${\rho}$ Fig.~\ref{sifig:rho_rhobar}(d))
Intense peaks in $\bar \rho({\theta, \phi})$ for $v_0 = 3.0$  in  \ref{sifig:rho_rhobar}(g) suggest that positional arrest of colloids in the ’solid phase’. 
The boundary between the lamellar and the lamellar with core states is determined by looking at the distribution of colloids ($\bar{\rho}$) in the region between 0-6 $\sigma$.  $\bar{\rho}$ calculated between $6-9 \sigma$ and $(\rho(r)$ are used to determined the crossover form the lamellar with core to the  solid structure. 

\begin{figure}[h]
\includegraphics[width=0.65\textwidth]{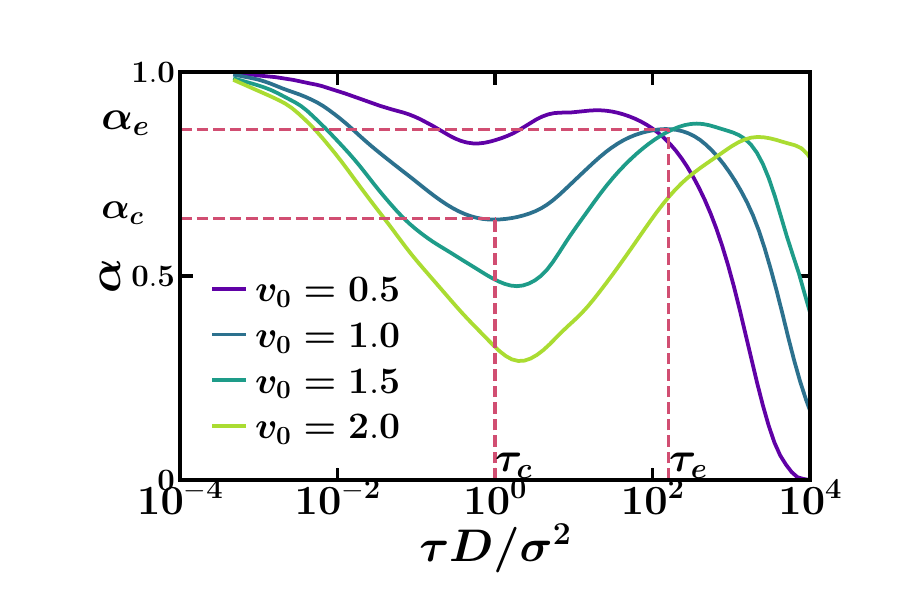}
\caption{The time varying exponent of anomalous diffusion, $\alpha$ for different values of $ v_0$. The exponent is close to unity at the earliest time scales. A plateau develops for all $v_0$ at nearly the same $\tau$, quantified by the minimum value $\alpha_c$ of $\alpha$. The escape time $\tau_e$ labelled in the plot is the local maxima at later time scales. $\tau_e$ increases with $v_0$ showing that the colloids are caged for longer duration of time as the cluster undergoes dynamical arrest.  }
\label{sifig:msd-alpha}
\end{figure}
%\clearpage
  
\section{ Estimating the caging and escaping time from the MSD}
Since the diffusion is anomalous at all time scales, the MSD can be written as $\msd(\tau) \propto \tau^{\alpha(\tau)}$, thus defining the time dependent exponent $\alpha(\tau)  \equiv \mbox{d}\log {\msd}(\tau)/ \mbox{d} \log \tau$. $\alpha$ changes with time following a general trend in all four states, see  Fig.~\ref{sifig:msd-alpha} -- $\alpha$ decreases from $1$ to a minimum value $\alpha_c$ at the caging timescale $\tau_c$, thereafter increasing to a maximum value $\alpha_e$ at the escape time $\tau_e$. $\tau_c \sim 1$ for all $v_0$; implying that colloids are trapped by neighbours when they have moved their own size. In contrast, escape from a cage is a collective manoeuvre and the related timescale  $\tau_{e}$  increases over three orders of magnitude from the fluid to the isotropic solid phase. 
    
\subsection{Individual trajectories of colloids}
    
Spatial heterogeneity in the dynamics of the colloids:  Fig.~\ref{sifig:particle-trajectory-msd} (a) shows a polar plot $r(t)$, $\phi(t)$ of
the radial and azimuthal components of the colloid position in spherical coordinates. Fig.~\ref{sifig:particle-trajectory-msd}(a-b) are different trajectories and their MSDs. The  Fig.~\ref{sifig:particle-trajectory-msd} show that dynamics of colloids strikingly vary with the initial location of the tagged particle in the cluster.
\begin{figure}
\includegraphics[width=0.65\textwidth]{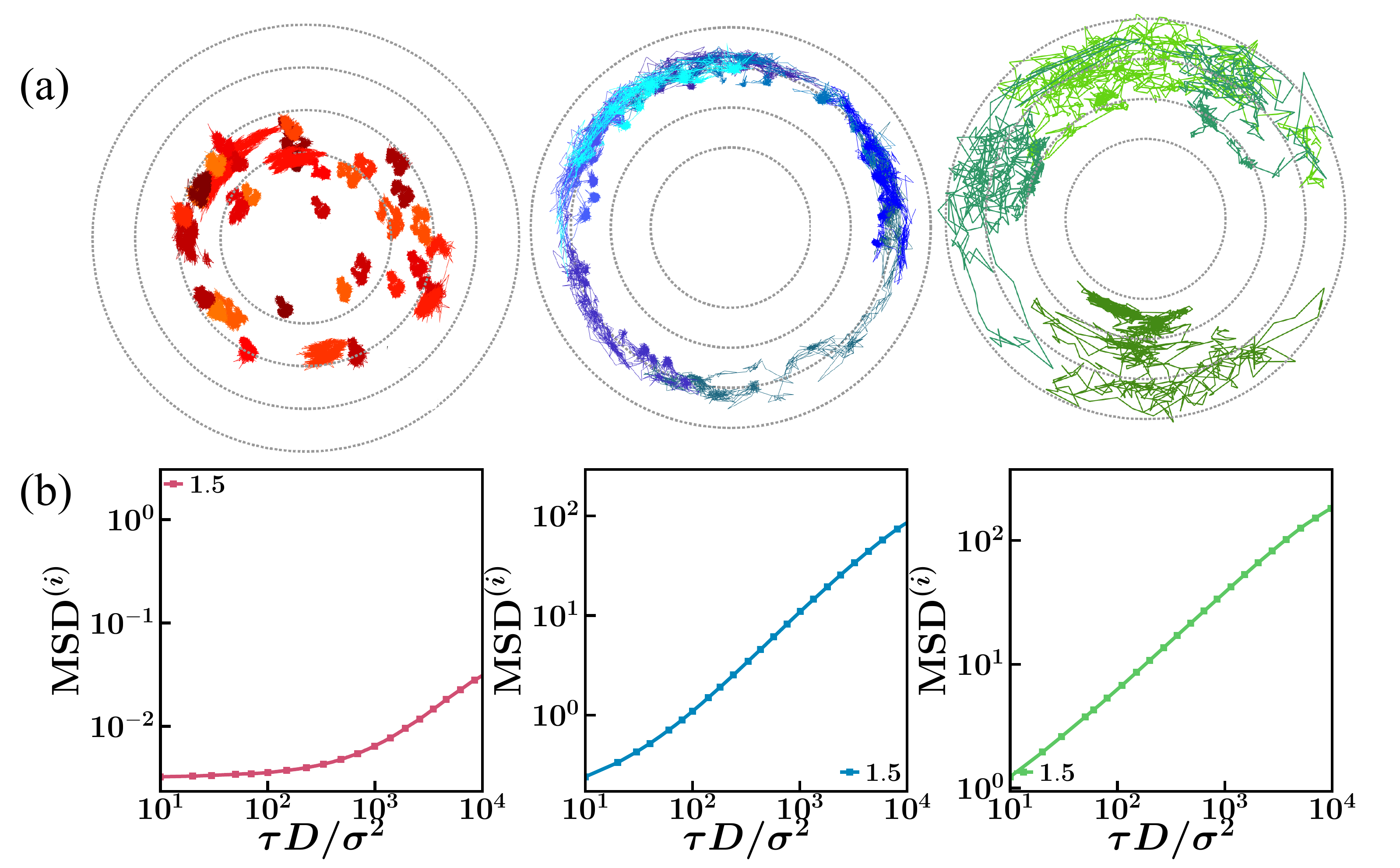}
\caption{(a) Various  trajectories showing different dynamics: trapped, intermittent caging and hoping, system spanning trajectories. (b) MSD of colloids for $v_0=1.5$, initially located in the core, corona and interface. }
\label{sifig:particle-trajectory-msd}
\end{figure}

\newpage
\section{Non-Gaussian fluctuations}  
In the main text we have shown that the self part of the Van Hove function assumes distinct forms as $v_0$ is varied. Particularly, it is interesting that the cage breaking dynamics becomes explicit in the inner dome and the wide tail of $G$. To quantify the distributions further we have fitted the curves to the $q-$Gaussian that is known to describe systems with long range correlations. The $q$-Gaussian of a length $x$ scaled by $\ell$ is $\left[ 1 - (1-q) (x/\ell)^{2} \right]^{\frac{1}{1-q}}$. Fig. \ref{sifig:particle-trajectory}(a-b) shows the results of the fits for values of $v_0$ specified in the figure. The fits have also been carried out for a large range of $\tau$ and by changing the total number $N$ of colloids. The observed trends are summarized in Fig. \ref{sifig:particle-trajectory}(c-f). 

\begin{figure}
\includegraphics[width=1\textwidth]{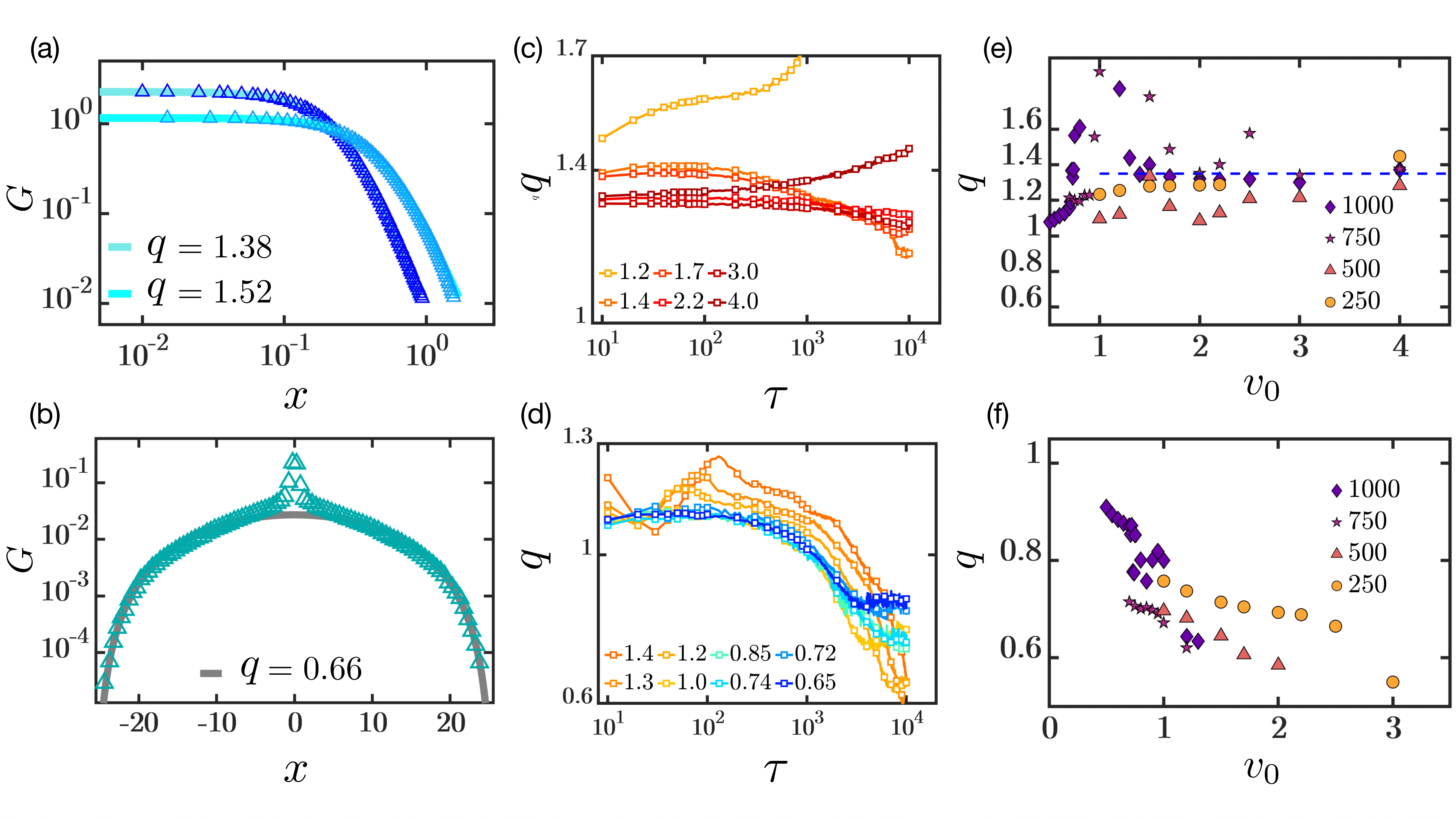}
\caption{Characterizing the non-Gaussian fluctuations: (a) the inner dome of $G$ for colloids in the corona are fitted to q-Gaussian distribution in the solid phase. The data from simulations is plotted with markers and the solid line shows the fitting. The light blue curves correspond to $v_0 = 3.0$, and the dark blue curves correspond to $v_0 = 1.4$. The values of $q$ obtained from the fitting  are shown in the legend. (b) Similarly, the outer dome of $G$ is fitted to q-Gaussian distribution in the `lamellar with core' phase with $v_0 = 1.2$, the value of $q$ obtained from the fitting is shown in the legend. (c-d) The change in the non-Gaussian parameter with time is shown panels (c) inner dome (d) tail. The corresponding values of $v_0$ are shown in the legend. Panels (e) and (f) show the universal aspects of
the exponent averaged over the times $10^3 - 10^4$, obtained by varying $v_0$ and $N$. We find that $q$  approaches 1.35 deep into the solid phase for all $N$ and at long times. % the exponent $q$.
}
\label{sifig:particle-trajectory}
\end{figure}

\newpage
\section{Waiting time distribution}
  \begin{figure}[t]
\includegraphics[width=0.5\textwidth]{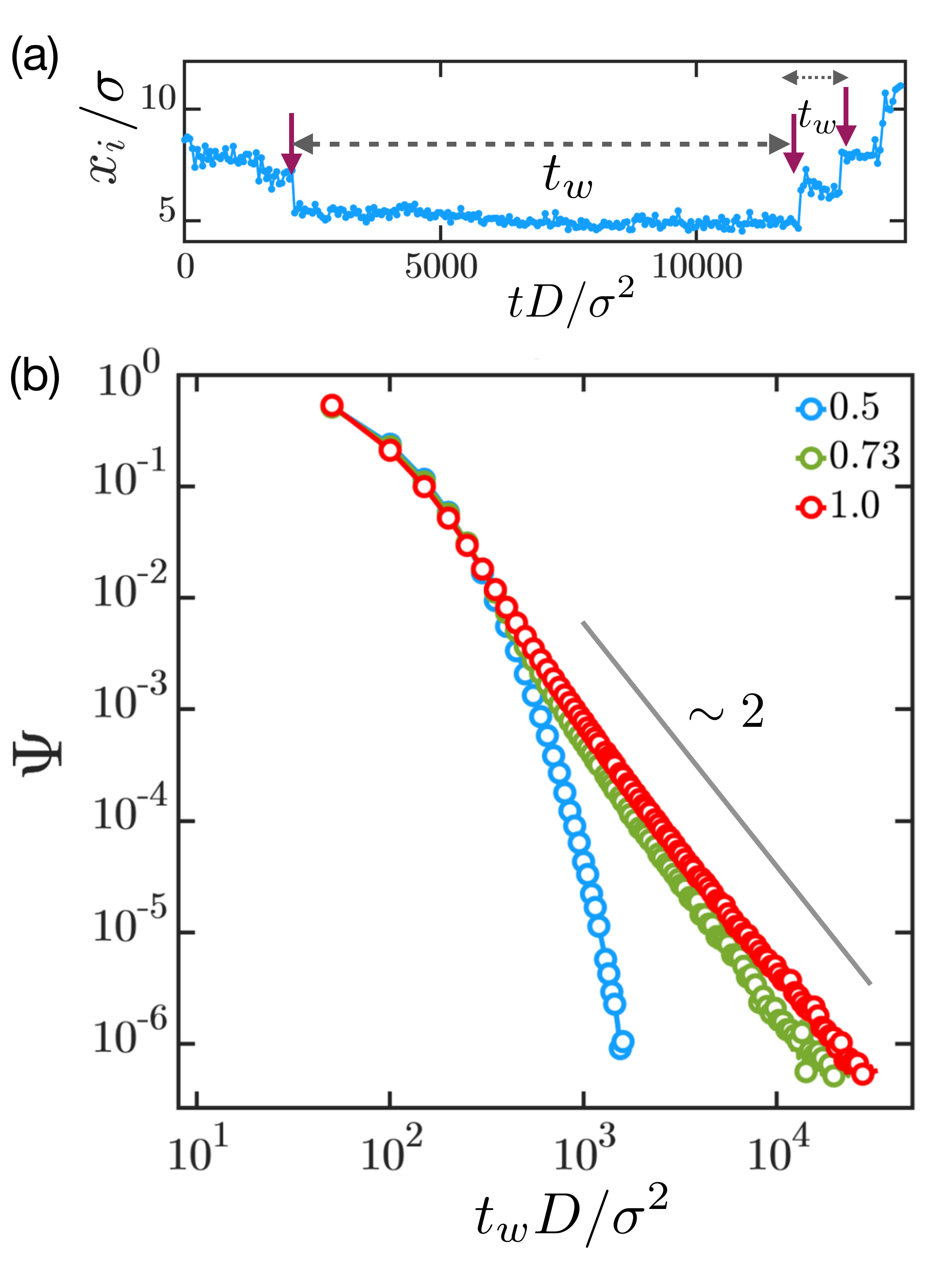}
    \caption{(a) Intermittent caging and hopping events. (b) The waiting time distribution goes from an exponential to power law.  The waiting time distribution is typically distributed as a power law with an exponent 2 for all $N$ in the solid phase.}
    \label{sifig:cage-hop}
    \end{figure}
The question that naturally arises form the description of the heterogeneous dynamics is: how frequent are the escapes from the caged states? As clear from the multitude of length-scales revealed by calculation of the Van-Hove functions, an accurate estimation of the length-scale for hops often possible in bulk systems (see \cite{Helfferich-2014}) is not possible here. We use a uniform definition for the jump event: a jump for particle $i$ at time $t$ is said to occur when $\Delta x_i=$ $|x_i(t+\tau)-x_i(t)|$ exceeds $ \sqrt{\text{MSD} (\tau)} $. The waiting time $t_w$ between two jumps is distributed as $\psi(t_w,\tau)$ and shown in Fig.~\ref{sifig:cage-hop}. The calculation reveals a striking feature for all system sizes, the distribution changes with $v_0$ from being close to an exponential distribution to a power law in the arrested state. A thorough understanding of the universality remains to be explored, however we can point out that with this definition of the jump subsumes several length-scales. Notice that the exponent $2$ is when the mean time between two hops diverges logarithmically. The appearance of these heavy tails point out that the system has developed memory; the amount of time a trapped colloid spends in its current cage depends on the amount of time it has spent in the cage already. 

\newpage
\section { Movies} 
\begin{enumerate}
  \item { SM1}:~{\tt S-Movie-1.mp4}, Trajectory of a few colloids in fluid phase $v_0=0.5$. Irrespective of the origin colloids traverse the entire droplet.
   \item  { SM2}:~{\tt  S-Movie-2.mp4}, Trajectory of a few colloids in lamellar with core phase $v_0=1.0$. Colloids at different regions of the droplet  exhibit varying dynamics.
\item { SM3}:~{\tt S-Movie-3.mp4}, Movie for fluid phase $v_0=0.5$, red bead shows a particle spanning entire droplet.
      \item { SM4}:~{\tt  S-Movie-4.mp4}, Movie for lamellar phase $v_0=0.7$,  red bead shows a particle spanning entire droplet. 
       \item{ SM5}:~{\tt S-Movie-5.mp4}, Movie for lamellar with core phase $v_0=0.8$,   Red bead shows caging and escape dynamics.      
             \item{ SM6}:~{\tt S-Movie-6.mp4}, Movie for lamellar with core phase $v_0=0.8$,  A slice of the droplet is shown. It shows immobile core and a dynamic outer region. Red bead clearly shows caging and escape or slow and fast dynamics.
        \item{ SM7}:~{\tt S-Movie-7.mp4}, Movie for solid phase $v_0=3.0$.            
\end{enumerate} 

   \end{document}